\documentclass[%
 reprint,
 amsmath,amssymb,
 aps,
 pra,
]{revtex4-2}

\usepackage[english]{babel}
\babelprovide[import=en]{en}
\usepackage{graphicx}% Include figure files
\usepackage{dcolumn}% Align table columns on decimal point
\usepackage{bm}% bold math
\usepackage{braket}
\usepackage{float}

\usepackage{color}

\begin{document}

%\preprint{APS/123-QED}

\title{Extended Wigner’s Friend Scenarios with Agent-like Observers \\on Quantum Computers}% Force line breaks with \\
%\thanks{A footnote to the article title}

\author{Joshua Laux}
\email{lauxj@ethz.ch} % change if ethz address will not be used in future
\affiliation{%
Institute for Theoretical Physics, ETH Zürich, 8093 Zürich, Switzerland
}%

\author{Eric G.~Cavalcanti}
\email{e.cavalcanti@griffith.edu.au}
\affiliation{%
Quantum and Advanced Technologies Research Institute, Griffith University, Gold Coast, QLD 4222, Australia
}%

% Add further authors here:
%
%\author{First Last}
%\affiliation{%
%Institute,
%University,
%City, Country
%}%

\date{\today}% It is always \today, today,
             %  but any date may be explicitly specified

\begin{abstract}
The Wigner's friend thought experiment raises the question of whether quantum theory can be applied consistently to systems that include observers. Extended Wigner's Friend Scenarios extend such scenarios to include spatially separate observers sharing entanglement. In the Local Friendliness (LF) framework, these lead to experimentally testable inequalities derived from the premises of Absoluteness of Observed Events and Local Agency -- a weaker set of premises than those required to derive a Bell inequality. This means that their violation has stronger implications, but LF tests are also experimentally more challenging, since they require coherent control of a quantum system that can serve as an observer.  

%Motivated by the ``Thoughtful'' version of the Local Friendliness no-go theorem–\cite{wiseman_thoughtful_2023}, which points towards future tests with human-level artificial agents implemented on quantum computers, 
Drawing from the literature on Artificial Intelligence, in this work we implement the ``friend'' with explicit agent-like functionality within a reversible quantum circuit. These agents store measurement outcomes, condition later operations on stored information, and, in the most structured case, use Born-rule probabilities to bet on the outcomes of their own future observations based on past observations. The circuits are simulated ideally and with an IBM-device noise model and executed on \texttt{ibm\_marrakesh}. Ideal simulations reproduce the maximal quantum violation of a Local Friendliness inequality up to finite-shot fluctuations, while hardware runs show positive LF violations for all implemented agents. 

These results provide a first step towards increasingly sophisticated agent-like observer models in LF experiments on quantum computers. By focusing on \emph{functionality} rather than on \emph{size} alone, this research program complements traditional tests of the limits of applicability of quantum theory in the macroscopic regime. Beyond the foundational implications, it also opens a new direction of research for the study of agents whose use of quantum theory is itself implemented within coherently controlled quantum dynamics.
\end{abstract}

%\keywords{Suggested keywords}%Use showkeys class option if keyword
                              %display desired
\maketitle

%\tableofcontents

\section{\label{sec:introduction}Introduction}

Quantum mechanics is one of the most successful theories for describing nature. Its predictions have been confirmed with remarkable precision. Despite its success, foundational puzzles such as the Wigner's friend thought experiment~\cite{wigner_remarks_1961}, the Einstein--Podolsky--Rosen (EPR) paradox~\cite{einstein_can_1935} and Bell's theorem~\cite{bell_einstein_1964} continue to raise fundamental questions about the interpretation of quantum mechanics.

The Wigner's friend thought experiment sharpens the quantum measurement problem~\cite{schlosshauer_decoherence_2005} by applying quantum mechanics to an isolated system involving an observer who performs a measurement on another quantum system. Applying the Schr\"odinger equation to this isolated system, an external observer will describe the friend and measured system as evolving unitarily into a superposition of quantum states assigning different outcomes to the friend's observation. On the other hand, applying the measurement postulate, the friend expects to obtain one or another definite outcome with probabilities given by the Born rule. This apparent conflict illustrates a fundamental tension at the heart of standard textbook formulations of quantum theory.
% This raises the question of whether measurement outcomes are absolute physical facts or may depend on the observer's perspective.

Extended Wigner's friend scenarios (EWFSs), developed in different forms by Brukner~\cite{brukner_no-go_2018} and Frauchiger and Renner~\cite{frauchiger_quantum_2018}, turn this tension into no-go results. Building on this line of work, Bong et al.~\cite{bong_strong_2020} showed that the premises of \textsc{Absoluteness of Observed Events} and \textsc{Local Agency} --- collectively denoted by ``Local Friendliness'' --- constrain the observable correlations through so-called LF inequalities. Quantum theory predicts violations of these inequalities, making the conflict between these premises and quantum predictions, in principle, experimentally testable. 

In contrast, Bell inequality derivations need an extra premise, e.g.~that the outcomes of all measurements are predetermined (by hidden variables)~\cite{cavalcanti_implications_2021}. This implies that violations of LF inequalities (in an LF scenario) have strictly stronger implications than the violations of Bell inequalities. The caveat is that an LF scenario requires an ``observer'' to play the role of the ``friend''. The proof-of-principle experiment of~\cite{bong_strong_2020} used simple photonic degrees of freedom as the ``friend'' (similarly to an earlier experiment~\cite{proietti_experimental_2019} testing Brukner's inequality). In these early experiments, one can maintain Local Friendliness by denying that the experiments involved a \emph{bona-fide} observer, raising the question of what counts as a conclusive LF test. 

Some interpretations, such as Relational Quantum Mechanics~\cite{rovelli_relational_1996} adopt a broad notion of observation, according to which any physical system undergoing an appropriate interaction --- including the simple photonic systems of the above experiments --- can be considered an ``observer''~\cite{di_biagio_stable_2021,di_biagio_relative_2025}. This has been challenged~\cite{brukner_qubits_2021}, illustrating that the interpretation of existing LF experiments depends crucially on what is accepted as a sufficient condition for observation.

This ``observer loophole'' was addressed in the ``Thoughtful Local Friendliness'' no-go theorem~\cite{wiseman_thoughtful_2023}, with suitably modified premises for which a conclusive experiment can be realised with the role of the ``friend'' played by a human-level artificial intelligence implemented on a quantum computer. However, although artificial intelligence has shown remarkable advances since that proposal, current quantum computers lag far behind. 

Although a decisive test may be far in the future, the fundamental implications motivate a research program to approach that goal, just as was the case for Bell's theorem. And just as for Bell's theorem, although one may assign a high credence that quantum predictions will hold, it may turn out to be impossible to violate the inequalities under ``loophole-free'' conditions, perhaps because new physics may come into play with observers, as Wigner originally thought~\cite{wigner_remarks_1961}. If this is the case, new physics could emerge somewhere along the path to this ultimate experiment. This motivates the development of tests with increasingly sophisticated ``observers''. In this vein, a recent work implements LF tests on quantum computers and characterizes their observer-like properties using figures of merit such as the branch factor~\cite{zeng_towards_2025}. 

In this work, we initiate a different direction for LF tests in quantum computers, aimed more directly towards the ultimate realisation with human-level AI, by introducing explicit \emph{agent-like functionality}. We implement simple agent-like systems as reversible quantum circuits and embed them into an LF scenario. The agents store measurement outcomes, act conditionally on stored information, and, in the most structured case, use Born-rule probabilities to choose a betting action. We study the resulting circuits in ideal simulations, realistic noise simulations, and on IBM quantum hardware, providing an intermediate step between elementary quantum systems and the much more sophisticated artificial observers envisioned in long-term LF proposals.

Beyond its foundational significance, this experimental program may also be relevant for quantum information theory. Implementing an agent as part of a quantum computation requires one to specify not only how information is stored and processed, but also how that information is used to assign probabilities and select subsequent actions. This raises a natural question of how probabilistic reasoning and decision-making should be implemented when the agent itself is embedded within the quantum system being described. As our results illustrate, this question is already relevant for simple agent-like quantum programs that can be implemented in today's quantum computers. Developing increasingly sophisticated agent-like programs may therefore provide a setting in which to study information processing for embedded quantum agents, to clarify how operational notions such as probability assignments, inference, and action can be implemented in such systems.

\section{\label{sec:local_friendliness}Local Friendliness and Extended Wigner's Friend Scenarios}

The minimal Extended Wigner's Friend Scenario considered here involves three observers: Alice, Charlie and Bob. In each run of the experiment a bipartite, entangled system is prepared and distributed to Alice's and Bob's side of the experiment. The subsystem $S_A$ is given to Charlie who sits inside the laboratory on Alice's side. Bob receives the subsystem $S_B$ and is able to measure this directly. Charlie, inside Alice's laboratory, performs a fixed measurement on $S_A$ and obtains an outcome $c$. Alice has a choice between two measurements labelled by $x\in\{1,2\}$ with outcomes $a$, while Bob has a choice between two measurements labelled by $y\in\{1,2\}$ with outcomes $b$. If Alice chooses $x=1$, she opens Charlie's laboratory, asks for Charlie's observed outcome and assigns her own outcome to be equal to Charlie's, that is $a=c$. If instead she chooses $x=2$, she performs a different measurement on the contents of Charlie's laboratory by reversing Charlie's measurement interaction and then measuring $S_A$ directly in a given basis. Bob, on the other side, performs one of his two possible measurements on $S_B$.

\begin{figure}[htbp]
    \centering

    \makebox[\columnwidth][l]{\textbf{(a)} \(x=1\)}
    \includegraphics[width=0.95\columnwidth]{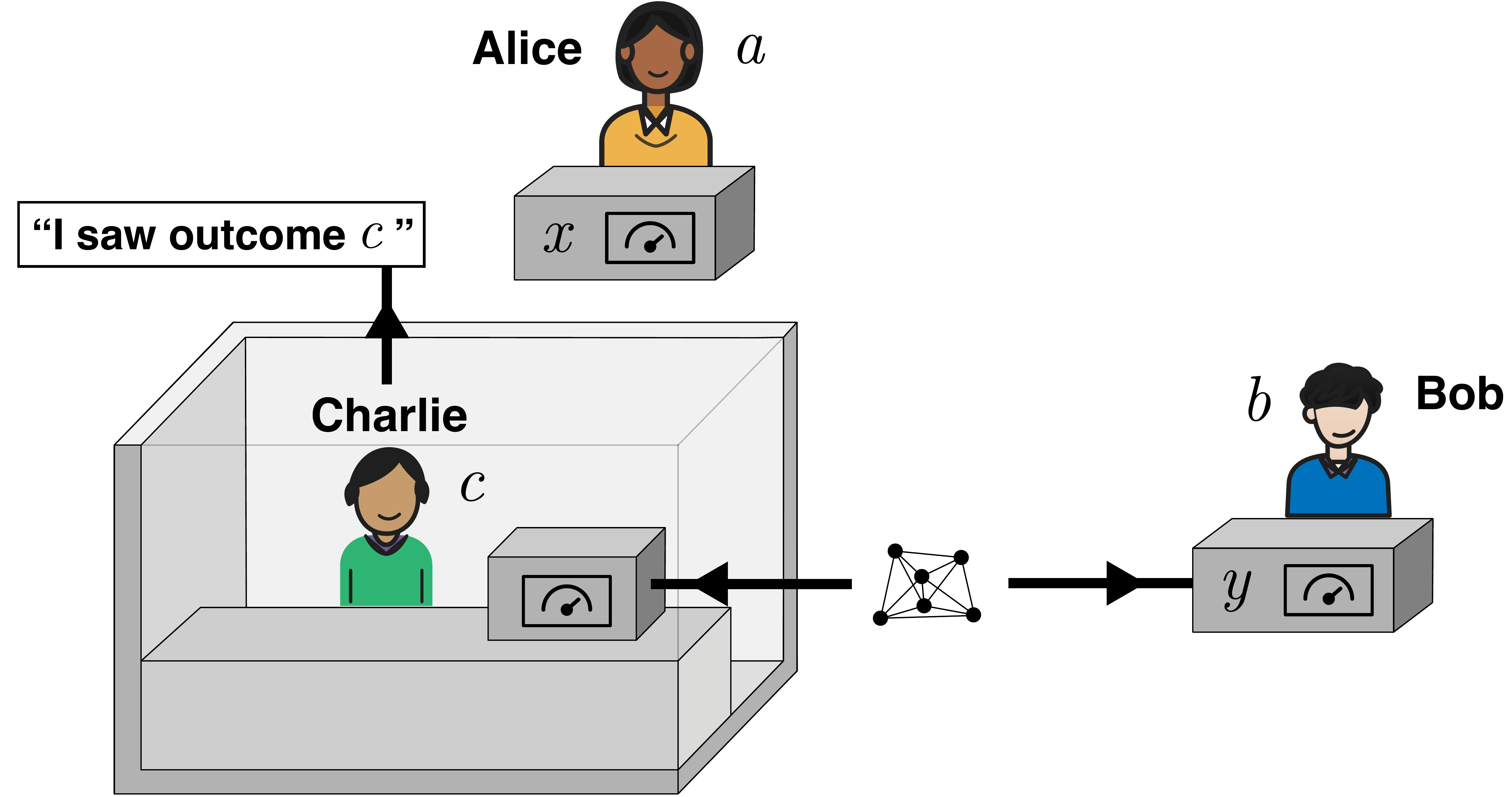}

    \vspace{0.2cm}

    \makebox[\columnwidth][l]{\textbf{(b)} \(x=2\)}
    \includegraphics[width=0.95\columnwidth]{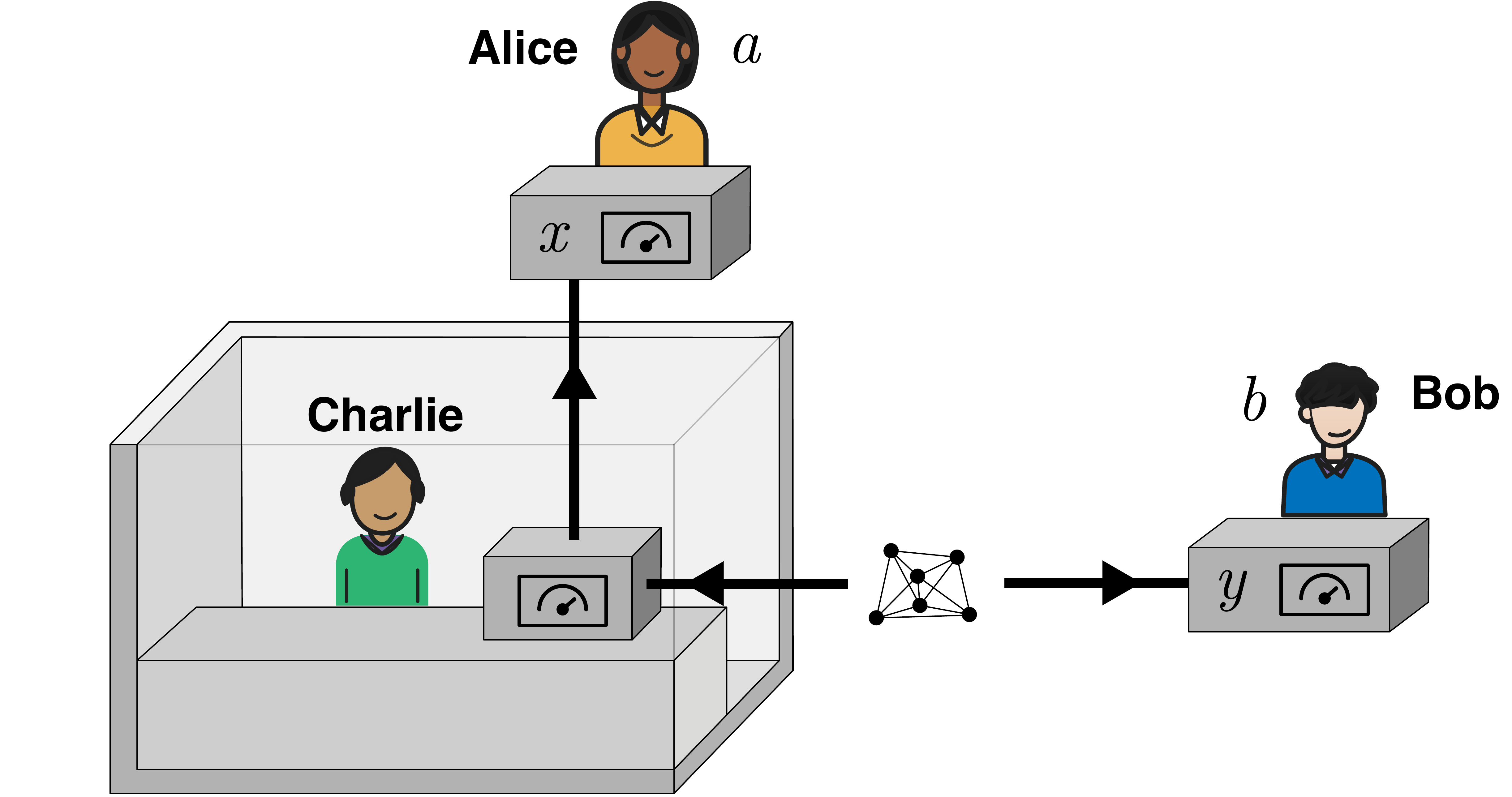}

    \caption{Extended Wigner's Friend Scenario with one friend, based on the setup discussed in Ref.~\cite{wiseman_thoughtful_2023}. Charlie is located inside Alice's laboratory and measures the subsystem \(S_A\), obtaining outcome \(c\), while Bob measures the entangled subsystem \(S_B\) directly. \textbf{(a)} For Alice's setting \(x=1\), Alice asks Charlie what outcome they observed and assigns her outcome \(a=c\). \textbf{(b)} For Alice's setting \(x=2\), Alice reverses Charlie's measurement interaction and restores the laboratory to the premeasurement state, and then measures \(S_A\) directly in a chosen basis. Bob independently chooses a setting \(y\) and obtains outcome \(b\).}
    \label{fig:EWFS_one_friend}
\end{figure}

This is the version of the EWFS used throughout the experiments presented here. It is minimal in the sense that it uses the least number of settings, outcomes, and friends while already sufficing for the violation of LF inequalities.

Similarly to Bell's theorem, extended Wigner's friend scenarios can be turned into experimentally testable no-go theorems by starting from metaphysical assumptions (by which we mean assumptions about physical theories) and deriving constraints on observable correlations. In this way, the Local Friendliness theorem~\cite{bong_strong_2020} follows the general strategy of Bell's theorem, adapted to scenarios in which observers themselves are included in the quantum description. The theorem shows that quantum correlations can violate constraints that follow from assumptions weaker than those used in Bell's theorem~\cite{cavalcanti_implications_2021,haddara_local_2025, haddara_properties_2025}.

In the original formulation by Bong et al.~\cite{bong_strong_2020}, ``Local Friendliness'' was defined as the conjunction of the premises of \textsc{No-Superdeterminism}, \textsc{Locality} and \textsc{Absoluteness of Observed Events}~\cite{bong_strong_2020}. Following the later reformulation by Cavalcanti and Wiseman~\cite{cavalcanti_implications_2021}, it is more convenient to express it in terms of the following two assumptions:
\begin{itemize}
    \item \textsc{Absoluteness of Observed Events}: Every observed event is an absolute single event.
    \item \textsc{Local Agency}: The only relevant events correlated with an intervention are in its future light cone.
\end{itemize}
Here, \textsc{Absoluteness of Observed Events} states that there is an absolute fact of the matter about any event observed by any observer. For example, if Alice observes her measurement device displaying an outcome ``spin up'', this is an absolute single fact: it is not a fact only relative to Alice, or only relative to a branch of the wavefunction, it is not the case that Alice also observed ``spin down'', etc. 

\textsc{Local Agency} states that freely chosen interventions are not correlated with relevant events outside their future light cone. The term ``intervention'', borrowed from the literature on causal models~\cite{pearl_causality_2000}, is used here rather than ``free choice'' to avoid associations with human free will~\cite{cavalcanti_implications_2021}. The qualifier ``relevant'' is important because an intervention may itself have physical causes in its past, such as the internal state of a device used to generate the setting. \textsc{Local Agency} therefore does not require interventions to be uncaused. Rather, an intervention refers to a choice of experiment made via suitable variables that share no common causes with, nor are caused by, other experimentally relevant variables.

These metaphysical assumptions constrain the observable correlations between Alice's and Bob's outcomes. More precisely, they restrict the empirical conditional probability distributions \(\wp(ab|xy)\) associated with outcomes \(a\) and \(b\) for measurement settings \(x\) and \(y\).

Under \textsc{Absoluteness of Observed Events}, Charlie's observed outcome \(c\) is also represented by an absolute event, so that the observable distribution admits an extension \(P(abc| xy)\) satisfying
\begin{equation}
\wp(ab| xy)
=
\sum_c P(abc| xy),
\qquad
\forall\,a,b,x,y.
\label{eq:aoe_marginal}
\end{equation}
For Alice's setting \(x=1\), she reads Charlie's recorded outcome. In the ideal case of perfect tracking this gives
\begin{equation}
P(a| c,x=1,y)
=
\delta_{a,c},
\qquad
\forall\,a,c,y.
\label{eq:aoe_tracking}
\end{equation}
This tracking condition is not a consequence of \textsc{Absoluteness of Observed Events} alone, but also assumes that Charlie's record is transmitted to Alice without error; its relaxation is considered in Sec.~\ref{sec:relaxed_lf}.

\textsc{Local Agency} further requires that the relevant events outside the future light cone of an intervention are uncorrelated with that intervention. For the present single-friend scenario, this implies
\begin{align}
P(c| xy) &= P(c),
&&\forall\,c,x,y,
\label{eq:la_c}\\
P(a| cxy) &= P(a| cx),
&&\forall\,a,c,x,y,
\label{eq:la_a}\\
P(b| cxy) &= P(b| cy),
&&\forall\,b,c,x,y.
\label{eq:la_b}
\end{align}
Combining these conditions, the observable correlations compatible with LF can be written as
\begin{equation}
\wp(ab| xy)
=
\begin{cases}
\displaystyle
\sum_c
\delta_{a,c}\,
P(b| cy)\,
P(c),
& x=1,\\[3mm]
\displaystyle
\sum_c
P_\mathrm{LA}(ab| cxy)\,
P(c),
& x\neq1,
\end{cases}
\label{eq:lf_model}
\end{equation}
where the subscript in \(P_\mathrm{LA}(ab| cxy)\) denotes that for each value of \(c\), the marginals of this conditional distribution must satisfy the \textsc{Local Agency} constraints in Eqs.~\eqref{eq:la_a} and~\eqref{eq:la_b}. Importantly, these conditions do not require the factorization
\(P(ab| cxy)=P(a| cx)P(b| cy)\), unlike Bell's assumption of Local Causality. This model defines the correlations from which the LF inequalities are obtained.

The set of observable distributions \(\wp(ab| xy)\) that admit a decomposition of the form in Eq.~\eqref{eq:lf_model}, subject to the tracking condition and the \textsc{Local Agency} constraints above, forms a convex polytope known as the LF polytope~\cite{haddara_local_2025}. As for Bell polytopes~\cite{brunner_bell_2014}, its vertices correspond to extremal behaviours, while its facets define linear inequalities satisfied by all LF-compatible correlations. An observed distribution lying outside this polytope is therefore incompatible with the conjunction of the LF assumptions and the tracking condition used in the model.

For the minimal one-friend scenario considered here, the LF polytope coincides with the corresponding Bell polytope~\cite{haddara_local_2025}. Consequently, the facet inequality used below has the same mathematical form as a Bell inequality, even though its interpretation in the present experiment is different: it is derived within an Extended Wigner's Friend Scenario and tests the LF assumptions associated with Charlie's observed event.

As is standard in CHSH-type scenarios, Alice's and Bob's outcomes are taken to be binary, \(a,b\in\{+1,-1\}\), with the corresponding observables denoted by \(A_x\) and \(B_y\). The inequality used in this work follows the convention used in Ref.~\cite{zeng_towards_2025} and is shifted such that the classical bound is zero:
\begin{equation}
S_{\mathrm{LF}}
=
-\langle A_1B_1\rangle
+\langle A_1B_2\rangle
-\langle A_2B_1\rangle
-\langle A_2B_2\rangle
-2
\leq 0 .
\label{eq:lf_inequality}
\end{equation}
More precisely, Eq.~\eqref{eq:lf_inequality} is equivalent to a Clauser--Horne--Shimony--Holt (CHSH) inequality~\cite{clauser_proposed_1969} under a relabelling of signs and settings. Accordingly, its classical bound is \(S_{\mathrm{LF}}\leq 0\), while quantum mechanics allows values up to
\[
S_{\mathrm{LF}}^{\max}=2\sqrt{2}-2,
\]
corresponding to Tsirelson's bound for CHSH~\cite{cirelson_quantum_1980}.

\subsection{\label{sec:relaxed_lf}Relaxed Local Friendliness}

The LF model above assumes perfect tracking between Charlie's recorded outcome and Alice's outcome for \(x=1\), as expressed in Eq.~\eqref{eq:aoe_tracking}. In a realistic experiment this condition may be imperfect because of errors in the physical implementation of Charlie's record and its subsequent readout by Alice, including gate noise, decoherence, and measurement errors. Following Moreno et al.~\cite{moreno_events_2022}, it can be relaxed to
\begin{equation}
P(a=c| x=1,y)\geq 1-\varepsilon
\qquad \forall\, y ,
\label{eq:relaxed_tracking}
\end{equation}
where \(\varepsilon=0\) recovers the ideal LF condition. Moreno et al. define the
corresponding set of correlations as the ``relaxed Local Friendliness'' set~\cite{moreno_events_2022}.

For the two-setting scenario considered here, the bound of
Eq.~\eqref{eq:lf_inequality} is shifted to
\begin{equation}
S_{\mathrm{LF}}\leq 4\varepsilon .
\label{eq:relaxed_lf_bound}
\end{equation}

For a fixed upper bound \(\varepsilon\) on the tracking error, a margin of more than three standard errors of the mean is obtained when
\begin{equation}
S_{\mathrm{obs}}-3\sigma_S>4\varepsilon,
\end{equation}
where \(S_{\mathrm{obs}}\) is the observed LF value and \(\sigma_S\) is its standard error. Equivalently,
\begin{equation}
\varepsilon<
\varepsilon_{\max}
=
\frac{S_{\mathrm{obs}}-3\sigma_S}{4}.
\label{eq:epsilon_max}
\end{equation}
We use \(\varepsilon_{\max}\) as a diagnostic threshold. Comparing it with an estimated tracking error does not by itself establish a \(3\sigma\) violation of the relaxed bound, since this comparison does not incorporate uncertainty in the tracking-error estimate.

\section{\label{sec:agent_observers}Agent-Like Observers}

Importantly, applying these inequalities to a particular experiment requires the additional assumption that the system playing the role of the friend qualifies as an ``observer'', or equivalently, that the process implements an ``observation'' of some physical quantity. That is, any experiment tests Local Friendliness in conjunction with some \emph{sufficient condition for observation}. For the early photonic experiments~\cite{proietti_experimental_2019, bong_strong_2020}, the condition can be described as: \emph{if via a suitable interaction a physical system $O$ obtains information about a variable $X$ of a system $S$, then $O$ has thereby observed $X$}~\cite{trotter_philosophical_2021}.

If one accepts this sufficient condition, then (modulo locality and detection loopholes), Local Friendliness has been experimentally violated. In order to maintain Local Friendliness, therefore, one must reject this sufficient condition, opening the question of what minimal sufficient condition for observation one would accept in its place. In the present work, we consider more stringent conditions, whereby a physical system qualifies as an observer if it can not only acquire information about another physical system but also use it to determine subsequent actions. That is, if the system functions as an \emph{agent}.

In the standard AI literature of Russell and Norvig~\cite{russell_artificial_2010}, an agent is defined as anything that can be viewed as perceiving its environment through sensors and acting upon that environment through actuators. The perceptual input at a given instant is called a percept and the complete history of everything the agent has ever perceived is called the percept sequence. Mathematically, the behaviour of an agent is described by an agent function that maps percept sequences to actions. In this sense, an agent is defined by a functional role: receiving information from the environment and producing actions on that basis.

%This gives a more structured notion of observer than simply saying that a system becomes correlated with another. %In a Wigner's friend scenario, the relevant features of an observer are that it receives information, stores or processes it internally and can act on the basis of it. 
Agent theory therefore provides a simple and systematic language for discussing more stringent conditions for observers in LF experiments. The aim is not to claim that such toy agents are already equivalent to human observers. Rather, they should be understood as a hierarchy of increasingly sophisticated and experimentally implementable observer models, that capture some of the informational and decision-like structure that seems relevant in Wigner's friend scenarios.

Related work has also emphasized the role of agency. Frauchiger and Renner consider agents who themselves use quantum theory to make predictions and reason about the predictions of other agents~\cite{frauchiger_quantum_2018}. Baumann and Brukner treat Wigner's friend as a rational agent who uses quantum theory to assign probabilities and make predictions~\cite{baumann_wigners_2020}. Cavalcanti similarly discusses the perspective of a quantum agent in terms of probability assignments and betting commitments~\cite{cavalcanti_view_2021}. Vilasini and Woods developed a general quantum-circuit framework for reasoning by agents in extended Wigner’s friend scenarios~\cite{VilasiniWoods2026}. These viewpoints motivate going beyond an observer that merely becomes correlated with a system and considering systems that store information, process it, and use it to determine later actions.

%\egc{I think the following descriptions should be at a much higher level of generality, just explaining in broad terms what these types of agents are, and what they will be doing in our implementation in particular. You can say that the betting agent will be represented as having a wallet register, from which the agent will take funds for the bet and into which they will receive any payoff. But avoid referring to specific details of the implementation for now. It is hard to follow as it is, and it feels like we are just going to repeat all of this again in the next section.} 

%\jl{Done. I made the introduction chapters of the agents much simpler, and put all technical explanations to the later sections where we describe how exactly the agents are implemented.}

We consider three simple agent models that differ in how they use the information obtained from an observation. Each model plays the role of Charlie in the EWFS described in Sec.~\ref{sec:local_friendliness}. Their reversible circuit implementations are described in Sec.~\ref{sec:reversible_implementation}.

\subsection{\label{sec:reflex_agent}Reflex Agent}

We begin with a simple reflex agent in the sense of Russell and Norvig~\cite{russell_artificial_2010}. Such an agent maps its current percept directly to an action according to a fixed condition--action rule, without using an internal model of the environment. In our implementation, the agent records a binary measurement outcome and uses it to control a two-state system in the environment. This system can be interpreted, for example, as a binary device that is switched on for outcome 1 and left off for outcome 0. Although the current percept is recorded in memory, the action depends only on this percept, rather than on a history of observations.

\subsection{\label{sec:guessing_agent}Guessing Agent}

The Guessing Agent uses an earlier observation to predict the outcome of a later one. In our implementation, the agent records a first measurement outcome and predicts the outcome of a second measurement, with a known transformation applied to the system between the two measurements. The prediction rule is specified in advance using the Born-rule probabilities for this transformation. The agent therefore records as its guess the highest conditional Born probability outcome. The guess which can subsequently be compared with the result of the second measurement. Thus, the first memory record is used to determine an action directed towards a future observation.

\subsection{\label{sec:betting_agent}Betting Agent}

We now consider a Betting Agent, which uses an earlier observation to select the stake for a bet on a later measurement outcome. Unlike the Guessing Agent, which records a predicted outcome, the Betting Agent always bets on the same event and varies the amount committed. The experimenter derives the policy classically from Born-rule probabilities and the game's rules. The circuit executes this fixed policy without calculating probabilities or utilities during the run. In the terminology of Russell and Norvig~\cite{russell_artificial_2010}, this provides a minimal example of a policy motivated by expected utility. In our implementation, the available funds are represented by an external wallet register, from which funds are taken to place the bet and into which any payout is received.

As a baseline for the Betting Agent, we also implement an Always-$3/4$ Agent. This agent participates in the same betting game, but always commits a fixed stake, independently of the first observed outcome. This comparison allows us to assess the effect of conditioning the bet on stored information.

\section{\label{sec:reversible_implementation}Reversible Implementation of Agent Programs}

The agent models introduced above are implemented as reversible quantum circuits. Throughout the circuit descriptions, \(S_A\) denotes the system qubit with which Charlie interacts. Charlie's memory registers are denoted by \(M\), or by \(M_1,M_2\) when two measurement outcomes are stored. Additional registers \(R\), \(G\), and \(W=(W_1,W_0)\) denote, respectively, the reflex register, the guess register, and the wallet register used in the betting agents. Here, ``environment'' denotes the agent's task environment within Alice's coherently controlled laboratory. The memory--action distinction is functional, with the policy implemented by the fixed gates coupling these registers.

Charlie's interaction with the system is modeled coherently. Charlie interacts with the system and records the computational-basis value of \(S_A\), which is modeled by a CNOT gate. From Charlie's perspective, this interaction represents the measurement of \(S_A\). From Alice's external description, the same process is represented as a reversible unitary interaction that correlates \(S_A\) with Charlie's memory register. This reversibility is essential for the EWFS: for Alice's setting \(x=2\), Charlie's unitary interaction is reversed before Alice measures \(S_A\) directly.

\subsection{\label{sec:reflex_circuit}Reflex Agent}

In the quantum circuit, Charlie's percept is obtained from the computational-basis value of the system qubit \(S_A\). This is implemented by a CNOT gate from \(S_A\) to \(M\), which records the value \(c\in\{0,1\}\) in Charlie's memory. The register \(M\) therefore carries the imprint of the value obtained in the interaction. Conditioned on this recorded value, Charlie then acts on the qubit \(R\) by preparing it in state \(\ket{0}\) for \(c=0\) and in state \(\ket{1}\) for \(c=1\). Hence, the agent's behaviour is fully specified by an immediate percept--action mapping. With this interpretation, Charlie is represented by the memory qubit \(M\), while the qubit \(R\) is part of the environment, as it is the target of the agent's action.

\begin{figure}[htbp]
    \centering
    \includegraphics[width=\columnwidth]{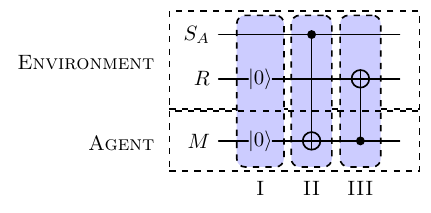}
    \caption{Quantum-circuit implementation of the simple Reflex Agent. \textbf{Step~\textsc{I}:} the memory qubit $M$ and the reflex qubit $R$ are initialized in state $\ket{0}$. \textbf{Step~\textsc{II}:} Charlie obtains the percept by correlating the memory qubit \(M\) with the computational-basis value \(c\) of \(S_A\), implemented by a CNOT gate from \(S_A\) to \(M\). \textbf{Step~\textsc{III}:} conditioned on $c$, Charlie acts on the qubit $R$ by setting it to the corresponding computational-basis state $\ket{0}$ or $\ket{1}$. In this interpretation, $S_A$ and $R$ belong to the environment, while $M$ represents the agent.}
    \label{fig:reflex_agent}
\end{figure}

\subsection{\label{sec:guessing_circuit}Guessing Agent}

A quantum-circuit implementation of the Guessing Agent uses two memory qubits \(M_1\) and \(M_2\), which store the first and second recorded outcomes, respectively. The register \(G\) stores Charlie's guess. Whether the guess was correct is evaluated only after the classical readout by comparing \(G\) with \(M_2\).

\begin{figure*}[htbp]
    \centering
    \includegraphics[width=0.80\textwidth]{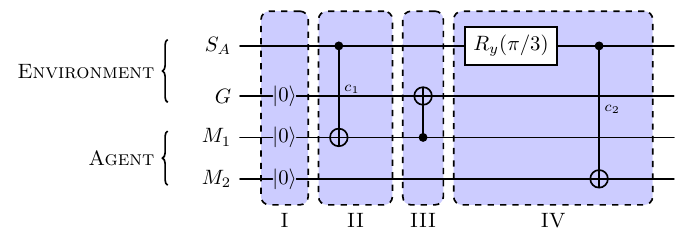}
    \caption{Quantum circuit for the Guessing Agent. \textbf{Step~\textsc{I}:} initialization of all registers. \textbf{Step~\textsc{II}:} Charlie records the first outcome $c_1$ in memory register $M_1$. \textbf{Step~\textsc{III}:} Charlie stores the guess in the register $G$. \textbf{Step~\textsc{IV}:} the system qubit is rotated by $R_y(\pi/3)$ and the second outcome $c_2$ is recorded in $M_2$. In this interpretation, $S_A$ and $G$ are treated as part of the environment, while $M_1$ and $M_2$ represent the internal memory of the agent.}
    \label{fig:guessing_agent}
\end{figure*}

The circuit, shown in Fig.~\ref{fig:guessing_agent}, proceeds in four steps. First, all registers are initialized. Next, the first recorded outcome \(c_1\) is stored in \(M_1\). Charlie's prediction is based on the known rotation \(R_y(\pi/3)\) applied to \(S_A\) between the two measurement interactions, where \(R_y(\theta)\) is defined in Eq.~\eqref{eq:Ry}. For this rotation, the Born rule gives
\begin{equation}
P(c_2=c_1|c_1)=\cos^2\left(\frac{\pi}{6}\right)=\frac{3}{4}.
\end{equation}
Charlie therefore writes the value \(c_1\) to the guess register \(G\). The system qubit is then rotated by \(R_y(\pi/3)\) and interacted with once more, with the second recorded outcome \(c_2\) stored in \(M_2\). After readout, the guess stored in \(G\) is compared with the second recorded outcome stored in \(M_2\).

With this interpretation, the system qubit \(S_A\) is part of the environment, since it is the system Charlie interacts with. The guess register \(G\) is treated as an external register on which Charlie writes the prediction, while the memory registers \(M_1\) and \(M_2\) represent the internal memory of the agent. Thus, the first memory record is not only stored, but is used by the agent to determine a later action.

\subsection{\label{sec:betting_circuit}Betting Agent}

Compared with the Guessing Agent, the Betting Agent requires an additional register to encode the available betting resource. The agent is modeled by two memory qubits, \(M_1\) and \(M_2\), representing the agent, together with an external two-qubit wallet register \(W=(W_1,W_0)\).

The betting game concerns whether the second computational-basis measurement of \(S_A\), after the same rotation \(R_y(\pi/3)\) used for the Guessing Agent, yields outcome 1. The corresponding conditional Born-rule probabilities are
\begin{equation}
P(c_2=1|c_1)=\begin{cases}\sin^2(\pi/6)=1/4, & c_1=0,\\ \cos^2(\pi/6)=3/4, & c_1=1.\end{cases}
\end{equation}
The payout is 1 if \(c_2=1\) and 0 otherwise. As a rule of the game, the minimum stake accepted by the ``bookie'' is set equal to the conditional probability of winning,
\begin{equation}
s_{\min}(c_1)=P(c_2=1|c_1). \label{eq:minimum_stake} 
\end{equation}
An admissible stake must therefore satisfy \(s\geq s_{\min}(c_1)\). For an accepted bet, the conditional expected net payoff is \(P(c_2=1|c_1)-s\). The Betting Agent maximizes this quantity by choosing the smallest admissible stake: \(1/4\) for \(c_1=0\) and \(3/4\) for \(c_1=1\). This policy is specified in advance and implemented by conditioning the wallet operation on the first recorded outcome.

The wallet encoding is \(w=W_1+W_0/2\), where \(W_0,W_1\in\{0,1\}\) are the computational-basis values of the wallet bits. At the beginning of the protocol, Charlie's wallet is initialized in the state \(\ket{01}\), corresponding to the value \(1/2\). Since the game is entered with a fixed bet of \(1/4\) already placed, Charlie initially still has another \(1/2\) available that may be chosen to be added to the bet.

\begin{figure*}[htbp]
    \centering
    \includegraphics[width=0.85\textwidth]{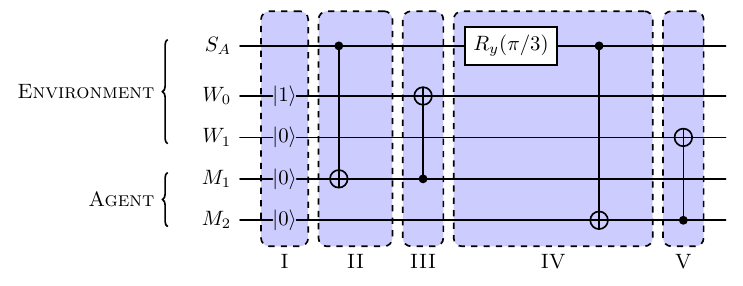}
    \caption{Quantum circuit for the Betting Agent. \textbf{Step~\textsc{I}:} initialization of all registers. \textbf{Step~\textsc{II}:} the Betting Agent records the first computational basis value $c_1$ in memory register $M_1$. \textbf{Step~\textsc{III}:} using this value, the wallet register is updated according to the chosen bet size. \textbf{Step~\textsc{IV}:} the system qubit is rotated by $R_y(\pi/3)$ and the second computational basis value $c_2$ is recorded in $M_2$. \textbf{Step~\textsc{V}:} the wallet is updated to encode the corresponding payoff. In this interpretation, the system qubit $S_A$ and the wallet register $W=(W_1,W_0)$ are treated as part of the environment, while $M_1$ and $M_2$ represent the internal memory of the agent.}
    \label{fig:betting_agent}
\end{figure*}

The circuit, shown in Fig.~\ref{fig:betting_agent}, proceeds in five steps. First, all registers are initialized. Next, the Betting Agent records the first value \(c_1\) in \(M_1\). Based on this stored value, the wallet is updated to encode whether the agent leaves the bet at \(1/4\) or increases it to \(3/4\). The system qubit is then rotated by \(R_y(\pi/3)\) and recorded once more, with the second computational-basis value \(c_2\) stored in \(M_2\). Finally, the wallet is updated again to encode the corresponding payoff.

In this implementation, the wallet register is treated as part of the environment, since it is the external register on which the Betting Agent's action is written. The memory registers \(M_1\) and \(M_2\) represent the internal memory of the agent. Thus, the first memory record is used not only to store an outcome, but to determine a utility-dependent action.

The Always-\(3/4\) Agent uses the same betting-game structure, but does not condition its action on the first observed outcome. Specifically, the Step~\textsc{III} stake-selection gate $\mathrm{CNOT}_{M_1\rightarrow W_0}$ is replaced by an unconditional $X$ gate on $W_0$, while the Step~\textsc{V} payout gate $\mathrm{CNOT}_{M_2\rightarrow W_1}$ is unchanged. Both the wallet-preparation and stake-selection $X$ gates are retained, and Alice's $x=2$ branch reverses the complete forward sequence, including both operations, before applying her measurement-basis rotation.

\section{\label{sec:ewfs_implementation}Extended Wigner's Friend Scenario with Agent-Like Observers}

The agent circuits introduced above describe the interaction between the agent, playing the role of Charlie, and the system qubit \(S_A\) inside Alice's laboratory. To obtain the full Extended Wigner's Friend Scenario, two further ingredients must be added.

First, the system qubit \(S_A\) is embedded into an entangled bipartite state together with Bob's system qubit \(S_B\). The basis of the scenario is an entangled pair of qubits, with one qubit \(S_A\) located in Alice's laboratory, which also contains Charlie, and the other qubit \(S_B\) located on Bob's side. The system qubits are prepared in the Bell state
\begin{equation}
\ket{\Phi^+}_{S_A S_B}
=
\frac{1}{\sqrt{2}}
\left(
\ket{00}_{S_A S_B}
+
\ket{11}_{S_A S_B}
\right).
\label{eq:bell_state}
\end{equation}
The qubit \(S_A\) is then passed to Charlie and processed according to the corresponding agent circuit, while \(S_B\) remains on Bob's side.

Alice and Bob can each choose between two different measurement settings, \(x\in\{1,2\}\) for Alice and \(y\in\{1,2\}\) for Bob. For \(x=1\), Alice asks Charlie about their outcome. In the circuit implementation, this is represented by a classical readout of the memory qubit in which Charlie's first recorded outcome is stored. For \(x=2\), Alice reverses Charlie's unitary evolution and then measures the system \(S_A\) directly in a chosen basis. Importantly, for the more complex agents, this requires reversing the complete agent evolution that occurred after Charlie's interaction with \(S_A\), rather than only the initial measurement interaction. Bob independently measures \(S_B\) in one of two chosen bases.

Changes of measurement basis are implemented by applying a rotation
\begin{equation}
R_y(\theta)
=
e^{-i\theta\sigma_y/2}
=
\begin{pmatrix}
\cos(\theta/2) & -\sin(\theta/2)\\
\sin(\theta/2) & \cos(\theta/2)
\end{pmatrix}
\label{eq:Ry}
\end{equation}
before computational-basis readout. The angles are denoted by \(\alpha\) for Alice's setting \(x=2\), and by \(\beta_1\) and \(\beta_2\) for Bob's settings \(y=1\) and \(y=2\), respectively.

The rotation angles used in the LF measurements were chosen to maximize the ideal violation of the LF inequality. The resulting choice used throughout the implementation is
\begin{equation}
\label{eq:optimal_angles}
\alpha=\frac{3\pi}{2},
\qquad
\beta_1=\frac{3\pi}{4},
\qquad
\beta_2=\frac{\pi}{4}.
\end{equation}
For these angles, the ideal noiseless value of the LF expression is
\begin{equation}
S_{\mathrm{LF}}^{\max}
=
2\sqrt{2}-2
\approx 0.828 .
\label{eq:lf_maximal_quantum_value}
\end{equation}
The derivation of the optimal measurement angles is given in Appendix~\ref{app:lf_details}.

Second, the free measurement-setting choices of Alice and Bob, also referred to as interventions, must be incorporated. For this purpose, two additional choice qubits \(A_C\) and \(B_C\) are introduced, one for Alice and one for Bob. Each is initialized in the state \(\ket{0}\), transformed into an equal superposition by a Hadamard gate, and then measured in the computational basis. The resulting measurement outcomes define the settings selected by Alice and Bob in a given run of the experiment. Their integration into the complete EWFS circuit is illustrated for the Betting Agent in Fig.~\ref{fig:compiled_betting_ewfs}.

This procedure implements the measurement choices as quantum-generated random bits in the ideal quantum description. For the LF experiment, however, the relevant requirement is that the choices of Alice and Bob are independent of physical variables that may also influence the experimental outcomes. In the present implementation, the settings are generated internally by measuring qubits on the same quantum processor, and therefore this independence is not certified.

The correlators of the LF inequality are defined for outcomes \(A_i,B_i\in\{+1,-1\}\), whereas the quantum-computer readouts are \(a_i,b_i\in\{0,1\}\). The measured values are therefore relabeled according to
\begin{equation}
0\mapsto +1,
\qquad
1\mapsto -1.
\end{equation}
The resulting outcomes are used to evaluate the correlators entering \(S_{\mathrm{LF}}\).

\section{\label{sec:quantum_computer_implementation}Quantum-Computer Implementation}

The EWFS circuits described above are implemented in \texttt{Qiskit}. For each agent, the complete experiment is compiled into a single circuit containing the preparation of the entangled system, Charlie's reversible agent dynamics, the generation of Alice's and Bob's measurement settings, and the corresponding setting-dependent measurements.

\begin{figure*}[t]
\centering
\includegraphics[width=0.95\textwidth]
{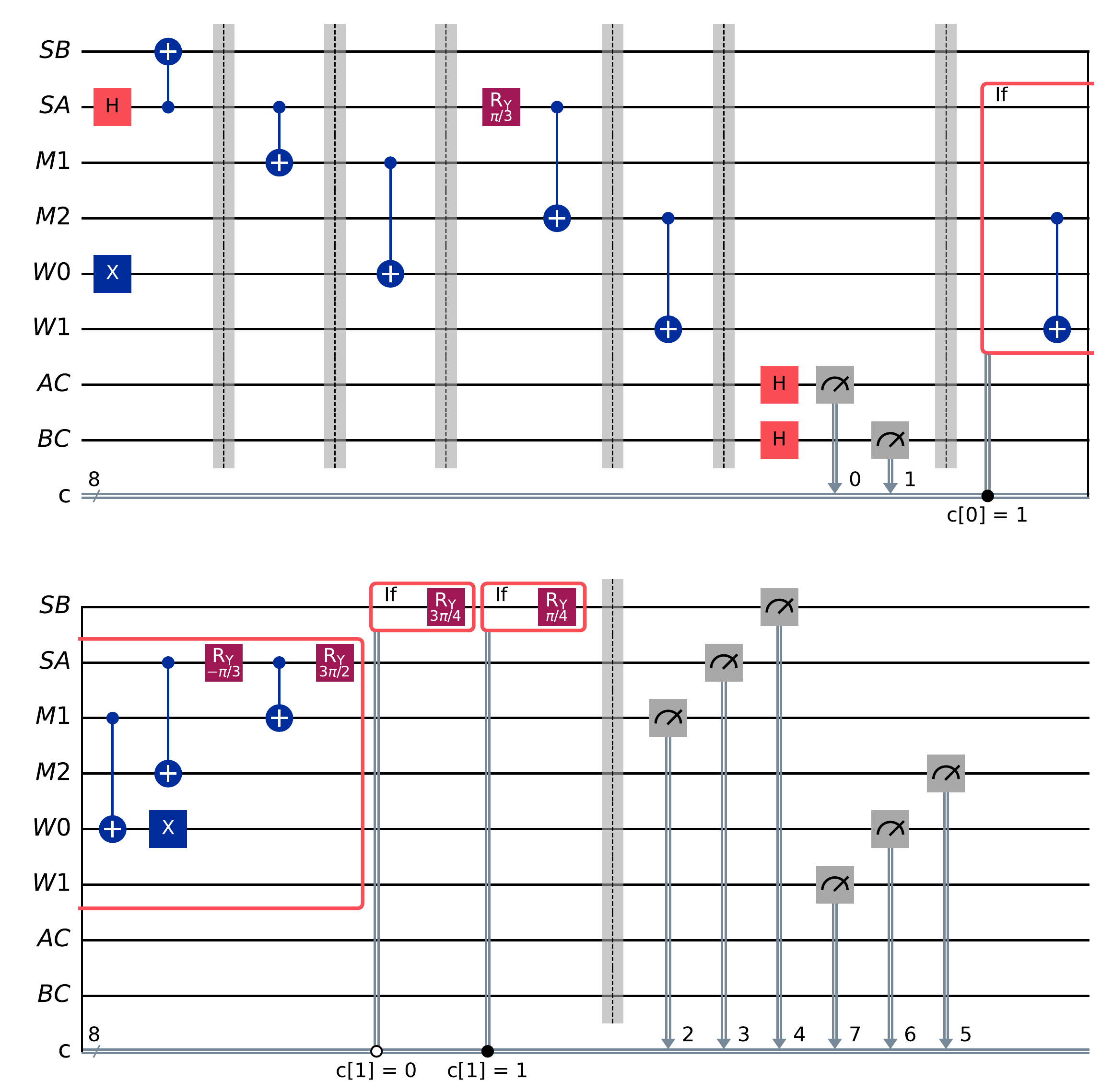}
\caption{Complete EWFS circuit for the Betting Agent. The Qiskit registers
\texttt{SA} and \texttt{SB} represent the system qubits \(S_A\) and \(S_B\),
while \texttt{M1} and \texttt{M2} represent Charlie's memory registers and
\texttt{W0} and \texttt{W1} represent the wallet register. The choice
registers \texttt{AC} and \texttt{BC} generate Alice's and Bob's settings.
From left to right, the circuit prepares the Bell state, implements Charlie's
reversible measurement and betting-agent dynamics, measures the choice
registers, applies the classically conditioned operations for Alice's and
Bob's settings, and finally reads out all registers.}
\label{fig:compiled_betting_ewfs}
\end{figure*}

A single circuit execution proceeds as follows. First, the system qubits \(S_A\) and \(S_B\) are prepared in the Bell state \(\ket{\Phi^+}\), after which \(S_A\) undergoes the reversible agent interaction representing Charlie's observation and subsequent information processing. The setting choices are generated after this interaction. Two additional qubits, \(A_C\) and \(B_C\), are initialized in \(\ket{0}\), transformed into equal superpositions using Hadamard gates, and measured in the computational basis. Their outcomes are stored as classical bits and determine the subsequent operations through Qiskit's classically conditioned control structure. The complete compiled circuit for the Betting Agent is shown in Fig.~\ref{fig:compiled_betting_ewfs} as a representative example of this implementation. The other agent circuits follow the same overall EWFS structure, differing in the reversible agent dynamics implemented between the Bell-state preparation and the setting-dependent measurements.

For Alice, one outcome of the choice qubit corresponds to \(x=1\), in which case her outcome is obtained from the memory register containing Charlie's first recorded outcome. The other corresponds to \(x=2\), for which the complete unitary evolution associated with Charlie's agent is reversed, after which \(R_y(\alpha)\) is applied to \(S_A\) and the system qubit is measured. For Bob, the value of his choice bit determines whether \(R_y(\beta_1)\) or \(R_y(\beta_2)\) is applied to \(S_B\) before computational-basis measurement. Thus, all four setting pairs \((x,y)\) occur within repeated executions of the same compiled circuit rather than being implemented as four separately constructed circuits.

Each shot returns a bit string containing the setting choices together with the relevant system and agent-register readouts. These bit strings are post-processed by conditioning on \(x\) and \(y\), extracting the corresponding outcomes \(a\) and \(b\), and evaluating the four correlators entering \(S_{\mathrm{LF}}\). All final readouts are computational-basis measurements, with the required measurement bases implemented through the corresponding \(R_y\) rotations before readout.

The choice qubits do not interact with the system or agent registers through two-qubit gates. On \texttt{ibm\_marrakesh}, they are mapped to physical qubits \(q_0\) and \(q_{155}\), respectively. This physical separation is an implementation choice and should not be interpreted as satisfying the space-like-separation requirement of a loophole-free LF experiment. Further details on the physical-qubit layout, native-gate decomposition, and transpilation are given in Appendix~\ref{app:hardware_implementation}. The complete circuit implementations are available in the accompanying code repository.

\subsection{\label{sec:simulations}Simulations}

Before running the circuits on quantum hardware, simulations are used to verify the ideal behavior of the implemented circuits and to estimate the effect of realistic device noise. We therefore perform both ideal simulations and noise simulations based on IBM calibration data.

Ideal simulations are performed with the \texttt{Qiskit AerSimulator} without a noise model. In this case, the circuits evolve according to the ideal unitary gates, while measurement outcomes are sampled according to the Born rule. For each circuit, \(10^5\) shots are used, so that the sampled outcome frequencies approximate the theoretical probabilities of the implemented circuit.

For the noise simulations, device-specific noise models provided by \texttt{Qiskit} are used in combination with the \texttt{AerSimulator}. These noise models are constructed from calibration data of IBM devices and include dominant noise sources such as gate errors, decoherence effects, and measurement errors. The simulations presented below use the noise model corresponding to \texttt{ibm\_marrakesh}. The noise simulations are executed with \(10^4\) shots, matching the number of shots used on hardware.

The noise model remains an approximation, and results obtained on hardware can deviate from the simulated values since additional effects may be present that are not included in the model.

\subsection{\label{sec:hardware_implementation}Hardware Implementation}

The implemented quantum circuits are executed on IBM quantum hardware via the \texttt{Qiskit} framework. The results presented in this work were obtained on \texttt{ibm\_marrakesh}. Before execution, the circuits are transpiled to match the native gate basis and connectivity of the device.

Since IBM devices are not fully connected, the circuit layout has to be chosen accordingly. If two qubits that need to interact are not directly connected, additional routing operations are required. Since these operations increase the circuit depth and therefore the sensitivity to noise, the circuits in this work are constructed such that no SWAP gates are needed.

For each agent, the logical qubits are mapped to physical qubits such that all required two-qubit connections are available directly. Using IBM calibration data for the main error sources, such as readout errors, CZ errors, and coherence times, a low-noise qubit placement is chosen. The logical-to-physical mappings and further implementation details are given in Appendix~\ref{app:hardware_implementation}.

The circuits are transpiled into the native gate set of the \texttt{ibm\_marrakesh} Heron architecture. The transpiled circuits are equivalent to the compiled logical circuits but expressed in the native gate set and physical-qubit layout of the selected backend. These represent the actual circuits executed in the experiment.

In contrast to the simulations, hardware executions are subject to the full physical noise present in the quantum processor. Deviations from the noise-simulation results can therefore arise from effects not captured by the noise model, such as crosstalk, calibration drift, spectator-qubit effects, and other device-specific imperfections.

The readout statistics from the simulations and hardware runs are post-processed to extract the empirical probabilities for each pair of settings. The correlators are evaluated as
\begin{equation}
\langle A_xB_y\rangle
=
\sum_{a,b=\pm1}ab\,P(a,b| x,y),
\end{equation}
and combined according to Eq.~\eqref{eq:lf_inequality}. The reported values are averaged over 10 independent runs and uncertainties denote the standard error of the mean.

\section{\label{sec:results}Results}

For the main EWFS experiments, the circuits were executed in 10 independent runs, using \(10^5\) shots per run for the ideal simulation, \(10^4\) for the noise simulation, and \(10^4\) on \texttt{ibm\_marrakesh} hardware. Uncertainties on the LF values and agent-performance estimates are reported as the standard error of the mean over these runs. The shot counts and uncertainty calculation for the separate tracking-error calibration are described in Appendix~\ref{app:transmission_circuits}.

\subsection{\label{sec:lf_results}Local Friendliness Violations}

\begin{figure*}[t]
    \centering
    \includegraphics[width=0.92\textwidth]
    {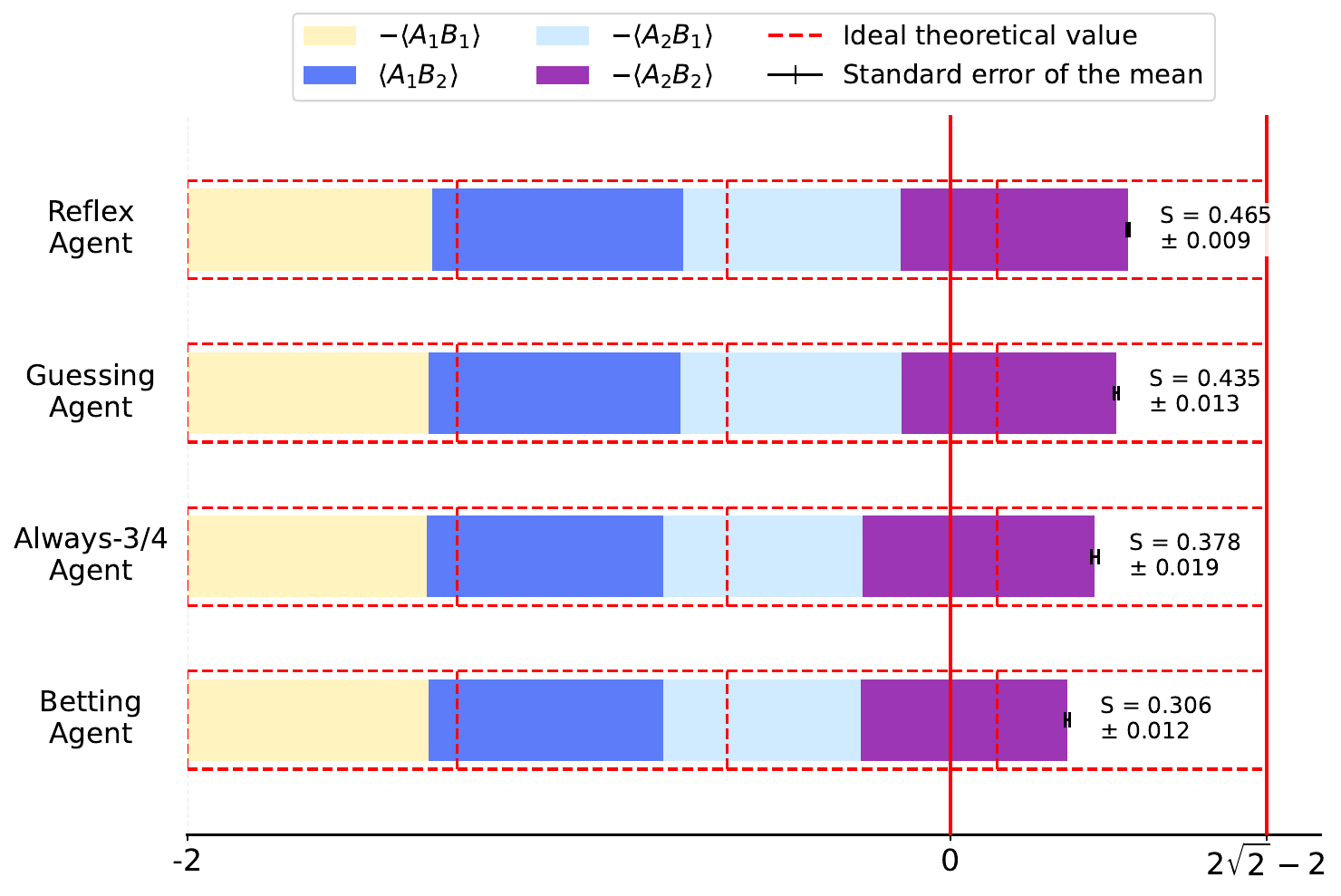}
    \caption{Observed LF violations from hardware runs on \texttt{ibm\_marrakesh} for the Reflex Agent, Guessing Agent, Always-$3/4$ Agent, and Betting Agent. The individual correlator terms entering the LF expression in Eq.~\eqref{eq:lf_inequality} are shown separately. The vertical red line at \(0\) marks the LF bound. The vertical red line at \(2\sqrt{2}-2\) marks the ideal quantum value for the chosen measurement angles. The red dashed lines indicate the ideal theoretical correlator values. The LF value for each agent is shown with the standard error of the mean over 10 runs, each consisting of \(10^4\) shots.}
    \label{fig:lf_violations_comparison}
\end{figure*}

An overview of the observed LF violations on \texttt{ibm\_marrakesh} hardware for the LF inequality in Eq.~\eqref{eq:lf_inequality} is shown in Fig.~\ref{fig:lf_violations_comparison} for the Reflex Agent, the Guessing Agent, the Always-$3/4$ Agent, and the Betting Agent.

As shown in Fig.~\ref{fig:lf_violations_comparison}, the hardware values of \(S_{\mathrm{LF}}\) remain positive by more than three standard errors of the mean for all four agents. The strongest violation is observed for the Reflex Agent. Including Bell-state preparation, the Reflex, Guessing, Always-\(3/4\), and Betting circuits execute respectively \(3,4,4,5\) CZ gates for Alice's setting \(x=1\), and \(5,7,7,9\) CZ gates for \(x=2\). The additional gates in the latter case implement the reversal of the agent evolution. These counts include gates inside the classically conditioned blocks. Gate counts alone do not determine the size of the violation: relevant factors also include circuit duration, which gates act on the qubits contributing to the LF correlators, and the gate and readout errors of the selected physical qubits.

Table~\ref{tab:lf_values} shows an overview of all agents' observed LF violations for the ideal simulation, the noise simulation, and the hardware runs on \texttt{ibm\_marrakesh}.

\begin{table*}[t]
\caption{LF violation values $S_{\mathrm{LF}}$ for all agents across ideal simulation, noise simulation, and hardware execution. Positive values indicate a violation of the LF bound. Uncertainties are the standard error of the mean over 10 independent runs.}
\label{tab:lf_values}
\begin{ruledtabular}
\begin{tabular}{lccc}
Agent & Ideal simulation & Noise simulation & \texttt{ibm\_marrakesh} \\
\hline
Reflex Agent
    & $0.826 \pm 0.003$
    & $0.679 \pm 0.010$
    & $0.465 \pm 0.009$ \\
Guessing Agent
    & $0.828 \pm 0.002$
    & $0.666 \pm 0.007$
    & $0.435 \pm 0.013$ \\
Always-$3/4$ Agent
    & $0.830 \pm 0.003$
    & $0.676 \pm 0.013$
    & $0.378 \pm 0.019$ \\
Betting Agent
    & $0.831 \pm 0.004$
    & $0.663 \pm 0.008$
    & $0.306 \pm 0.012$ \\
\end{tabular}
\end{ruledtabular}
\end{table*}

The ideal simulations reproduce the theoretical value \(2\sqrt{2}-2\simeq0.828\) for all four agents up to finite-shot fluctuations. The noise simulations reduce the violations, while the hardware values are reduced further. Nevertheless, all hardware values remain clearly above the LF bound.

The reduction is strongest for the more complex branches involving \(A_2\), where the agent evolution must be reversed before Alice measures the system. The Betting Agent shows the smallest hardware violation. This is consistent with the Betting Agent having the largest transpiled circuit depth and the largest number of entangling gates. In particular, the Betting Agent includes additional wallet qubits and conditional wallet updates. Although these wallet qubits are not themselves part of the LF correlator, they interact with the measured system and memory qubits before the final measurements.

\subsection{\label{sec:agent_evaluations}Agent Evaluations}

The different agents are also evaluated in terms of their intended behaviors and whether the circuit implementations reproduce them. The agents are evaluated using the same data as for the calculation of the LF violations. All agent-performance accuracies and mean payoffs reported in this subsection are evaluated using only shots with Alice's setting \(x=1\), combining both values of Bob's setting \(y\). Shots with \(x=2\) are excluded because Alice reverses the agent's evolution, so the final registers do not represent a completed action or bet. Estimates are calculated separately for each run and then averaged with equal weight across runs.

For the Reflex Agent, the readouts of the memory qubit \(M\) and the reflex qubit \(R\) should agree by construction. The data show perfect agreement for the ideal simulation. In the noise simulation, the reflex accuracy is \(98.7\%\pm0.1\%\), while for the hardware runs it is \(95.8\%\pm0.1\%\).

For the Guessing Agent, the relevant accuracy is \(P(G=M_2 | x=1)\), which is expected to be \(75\%\) under ideal conditions. The observed accuracy is \(74.3\%\pm0.2\%\) for the noise simulation and \(71.5\%\pm0.2\%\) for the hardware runs. These values provide strong evidence that the relevant behavior is captured correctly by the implementation.
Since these accuracies are inferred from noisy final readouts, they do not directly give the corresponding pre-readout accuracies. Under the assumption that the relevant noise processes do not artificially increase the observed agreement, they can be interpreted as lower-bound estimates of the implemented behavior.

For the Always-$3/4$ Agent, the intended behavior is to always place the \(3/4\) bet, independent of any previous measurement outcome. In the ideal simulation this behavior is reproduced perfectly. For the noise simulation, the observed accuracy is \(0.9945\pm0.0003\), while for the hardware runs it is \(0.9964\pm0.0003\).

As expected, the Betting Agent places its bets according to the Born Rule probabilities with perfect accuracy in the ideal simulation. The values found for
\(P(\text{bet }1/4| c_1=0, x=1)\) are \(98.6\%\pm0.1\%\) for the noise simulation and \(91.8\%\pm0.3\%\) for the hardware runs. For
\(P(\text{bet }3/4| c_1=1, x=1)\), we find \(98.7\%\pm0.1\%\) for both the noise simulation and the hardware runs.

The Betting Agent is also compared with the Always-$3/4$ Agent, which participates in the same betting game. For each run \(r\), the net payoff inferred from the wallet is 
\begin{equation} 
\widehat{\Pi}_r=\frac{1}{N_{r,1}}\sum_{i:x_i=1}\left(W_{1,i}+\frac12 W_{0,i}-\frac34\right), 
\label{eq:wallet_payoff_estimator} 
\end{equation}
where \(N_{r,1}\) is the number of shots with \(x=1\) in that run, including both Bob settings. The subtraction of \(3/4\) accounts for the initial wallet value \(1/2\) and the already-placed stake \(1/4\). The reported estimator is \(\widehat{\Pi}=R^{-1}\sum_{r=1}^R\widehat{\Pi}_r\), with its standard error calculated across the \(R=10\) run estimates.

\begin{figure}[htbp]
    \centering
    \includegraphics[width=\columnwidth]
    {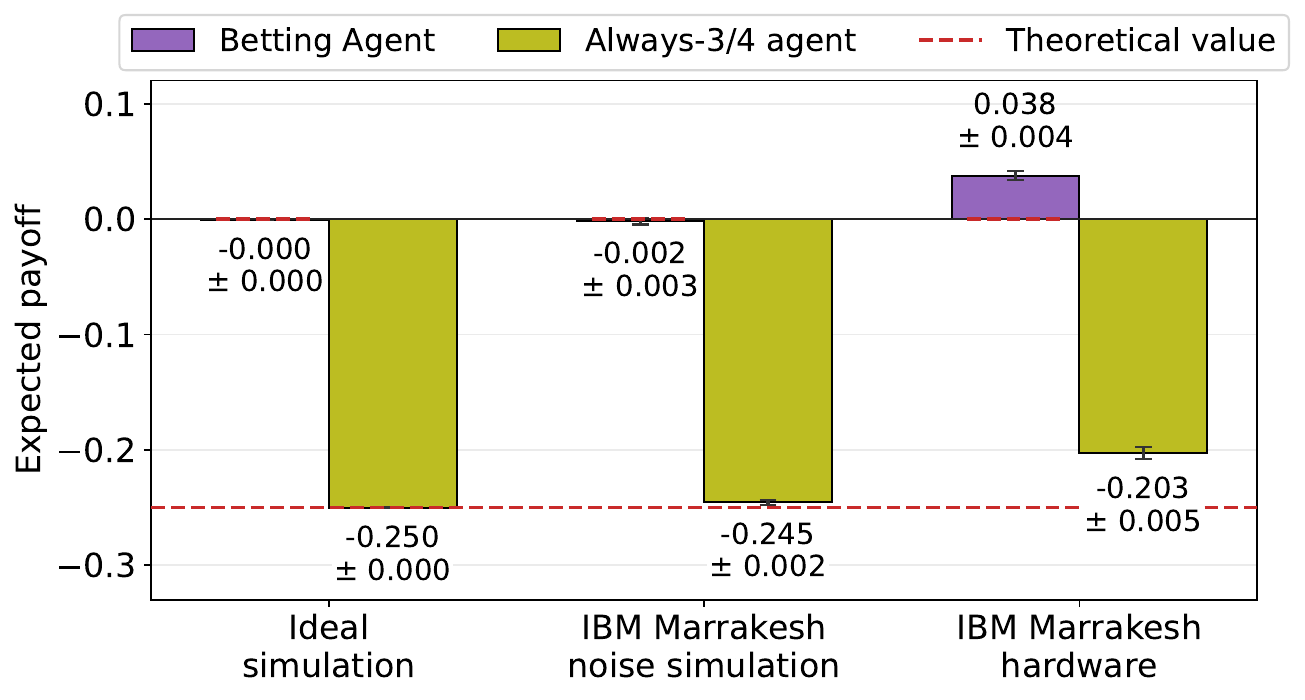}
    \caption{Comparison of the net payoff inferred from the final wallet readout relative to the initial funds over all runs of the Betting Agent and the Always-$3/4$ Agent participating in the same betting game. The plot shows the data for the ideal simulation, the noise simulation, and the hardware runs. Mean payoffs use Eq.~\eqref{eq:wallet_payoff_estimator}, retaining all \(x=1\) shots and combining both of Bob's settings; no admissibility or payout-consistency filter is applied.}
    \label{fig:betting_comparison}
\end{figure}

All \(x=1\) shots are retained, including readouts indicating a stake below \(s_{\min}(c_1)\) or a disagreement between the payout bit \(W_1\) and the outcome bit \(M_2\). The game's admissibility rule is therefore not enforced in this scoring: the reported quantity is a wallet-readout diagnostic. The comparison is shown in Fig.~\ref{fig:betting_comparison}.

Comparing the Betting Agent with the Always-\(3/4\) Agent confirms the expected difference in their mean net payoffs. Under ideal conditions, the unconditional probability of \(c_2=1\) is \(1/2\). The Always-\(3/4\) Agent therefore has expected net payoff \(1/2-3/4=-1/4\). The Betting Agent instead chooses the stake \(s=P(c_2=1| c_1)\), giving zero expected net payoff for either value of \(c_1\). The higher observed mean payoff of the Betting Agent is therefore consistent with the conditional policy derived from the Born rule and the imposed minimum-stake rule in Eq.~\eqref{eq:minimum_stake}. The reported payoffs are inferred from noisy final wallet readouts. On hardware, the payout bit $W_1$ disagrees with the recorded outcome $M_2$ in approximately $18\%$ of the $x=1$ shots for both agents. Replacing $W_1$ with $M_2$ in Eq.~\eqref{eq:wallet_payoff_estimator}, while retaining the same shots and averaging procedure, gives mean net payoffs of $-0.0290\pm0.0023$ for the Betting Agent and $-0.2691\pm0.0022$ for the Always-$3/4$ Agent, with uncertainties given by the standard error across runs. Thus, the Betting Agent's higher mean payoff persists under this alternative diagnostic, but its positive wallet-derived payoff should not be interpreted as exceeding the ideal game's optimal expected payoff. Since both registers are subject to noise, their disagreement does not identify the underlying error mechanism or establish which readout more accurately represents the outcome before readout.

\subsection{\label{sec:relaxed_lf_results}Robustness against Relaxed Local Friendliness}

To assess the robustness of the observed LF violations against imperfect tracking, we estimate the relaxation parameter \(\varepsilon\) introduced in Eq.~\eqref{eq:relaxed_tracking}. The target tracking-error probabilities are
\begin{equation}
\varepsilon_y=P(c\neq a| x=1,y).
\end{equation}
We estimate these probabilities using separate calibration circuits with fixed memory preparations, as described in Appendix~\ref{app:transmission_circuits}. Denoting the resulting estimates by \(\widehat{\varepsilon}_y\), we define
\begin{equation}
\varepsilon_{\mathrm{est}}
=
\max_{y\in\{1,2\}}\widehat{\varepsilon}_y.
\end{equation}
This maximum covers both estimated calibration errors, but does not establish an upper bound on the true tracking errors in the original experiment.

The corresponding estimated relaxed LF bound is \(4\varepsilon_{\mathrm{est}}\). For these calibration circuits, the tracking error vanishes exactly in the ideal noiseless simulation. Nonzero errors in the noise simulations and hardware runs can arise from preparation errors, gate noise, decoherence, and readout errors. The calibration therefore provides a practical estimate of \(\varepsilon\), subject to the assumptions described in Appendix~\ref{app:transmission_circuits}.

\begin{figure}[htbp]
    \centering
    \includegraphics[width=\columnwidth]
    {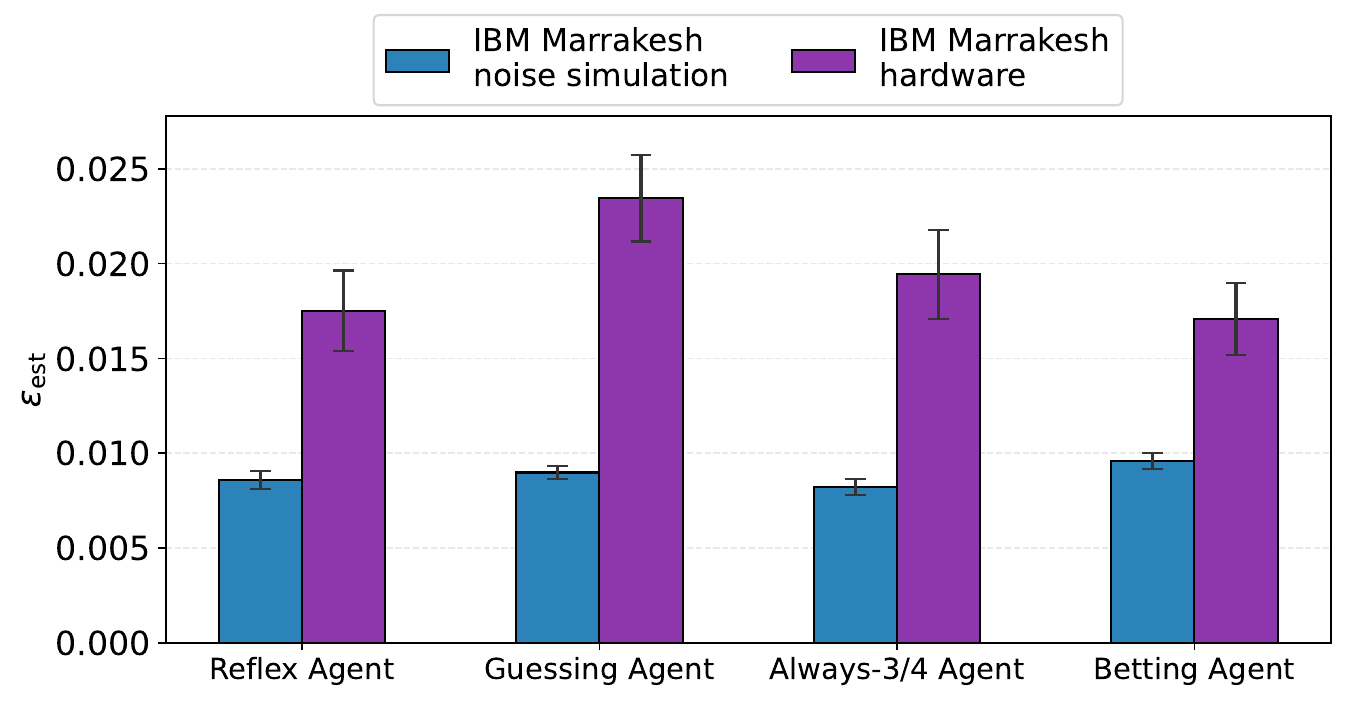}
    \caption{Calibration-based estimates of the tracking error. For each agent and each value of \(y\), the memory-readout error is averaged over runs and over the two fixed memory preparations. The reported value \(\varepsilon_{\mathrm{est}}\) is the larger of the resulting estimates for \(y=1\) and \(y=2\). Here, \(c\) denotes the intended memory-preparation label, and \(a\) is Alice's final readout. Only shots with \(x=1\) are included. Error bars show propagated standard errors for the selected fixed value of \(y\); they do not account for uncertainty associated with selecting the maximum.}
    \label{fig:tracking_values}
\end{figure}

Table~\ref{tab:epsilon_comparison} compares the hardware calibration estimates \(\varepsilon_{\mathrm{est}}\) with the thresholds \(\varepsilon_{\max}\) defined in Eq.~\eqref{eq:epsilon_max}. These thresholds account for the standard error of the observed LF value, but do not incorporate uncertainty in the calibration estimates.

\begin{table}[htbp]
\caption{Comparison of the hardware calibration estimates \(\varepsilon_{\mathrm{est}}\) with the thresholds \(\varepsilon_{\max}\) defined in Eq.~\eqref{eq:epsilon_max}. The quoted uncertainties are propagated standard errors for the selected fixed value of \(y\); they do not account for uncertainty associated with selecting the maximum. The calculation is described in Appendix~\ref{app:transmission_circuits}.}
\label{tab:epsilon_comparison}
\begin{ruledtabular}
\begin{tabular}{lcc}
Agent & $\varepsilon_{\mathrm{est}}$ & $\varepsilon_{\max}$ \\
\hline
Reflex
    & $0.0175 \pm 0.0021$
    & $0.1097$ \\
Guessing
    & $0.0234 \pm 0.0023$
    & $0.0992$ \\
Always-$3/4$
    & $0.0194 \pm 0.0023$
    & $0.0803$ \\
Betting
    & $0.0171 \pm 0.0019$
    & $0.0673$ \\
\end{tabular}
\end{ruledtabular}
\end{table}

For all four agents, the calibration estimate \(\varepsilon_{\mathrm{est}}\) lies below the corresponding threshold \(\varepsilon_{\max}\). Thus, the observed LF values remain above the estimated relaxed bounds with a margin exceeding three standard errors of \(S_{\mathrm{obs}}\). This comparison is conditional on the calibration assumptions and does not incorporate uncertainty in \(\varepsilon_{\mathrm{est}}\).

\section{\label{sec:discussion}Discussion}

The positive LF violation values obtained for all implemented agents show that LF inequalities can be violated using simple agent-like reversible circuit models on quantum computers in the one-friend scenario considered here. Compared with photon-path degrees of freedom as friends, the implemented agents provide a first step towards more structured observer-like systems. They do not constitute human-level artificial intelligence, but they implement limited observer-like properties: storing an outcome, conditionally acting on this information, and applying simple decision rules to determine subsequent actions. The Betting Agent implements a rudimentary use of quantum-theoretic predictions through a precomputed policy derived from conditional Born-rule probabilities.

The closest comparison is the quantum-computer implementation of Zeng et al.~\cite{zeng_towards_2025}, where the friend is made more observer-like by increasing a figure of merit called ``branch factor''. The present work follows a complementary direction. Rather than increasing the complexity of the branches as a target, it adds elementary agent functionality to the friend.  
A human-level artificial intelligence of the kind envisioned in Thoughtful Local Friendliness~\cite{wiseman_thoughtful_2023} would be expected to lie in a regime of very large branch factor, although its precise value is not known~\cite{zeng_towards_2025}. 

Ideal simulations reproduce the theoretical maximum value \(2\sqrt{2}-2\) for all implemented agents, showing that within the ideal quantum description the additional agent-like functionality does not itself reduce the LF violation. The noise simulations predict smaller but comparable violations across the four agents, whereas the hardware values decrease as the agent circuits become more complex. This trend is consistent with the increasing circuit depth, number of entangling gates, and exposure to hardware noise. Moreover, the growing discrepancy between the noise-model predictions and the hardware results indicates that the employed noise model does not capture all relevant device effects. Since increased agent functionality is necessarily accompanied here by increased circuit complexity, however, the present experiment cannot disentangle these two effects. The observed reduction can therefore be explained by ordinary hardware imperfections, but the data do not independently exclude an additional dependence on the physical complexity of the agent. All four agents nevertheless retain positive LF violations on hardware.

For Relaxed LF, the tracking error was estimated using modified circuits with fixed memory preparations. The observed LF values exceed the corresponding estimated relaxed bounds for all four agents. However, the calibration procedure does not establish a rigorous upper bound on the tracking error in the original experiment, and the comparison with \(\varepsilon_{\max}\) does not incorporate uncertainty in the calibration estimate. These results therefore support robustness against tracking errors within the assumptions of the applied estimation method.

The experiment is not loophole-free. In the intended EWFS, Alice and Bob must be space-like separated. Using the scheduling timing data of the hardware runs, the required separation between Alice and Bob is of order \(1.8\,\mathrm{km}\) for the longest circuit. Since all qubits in the present implementation are located on the same IBM device, this locality condition is not satisfied. A possible future implementation would require spatially separated devices sharing an entangled qubit pair. For the one-friend scenario implemented here, Alice's device would have to be a quantum computer with a quantum input, while Bob's could be a simpler measurement device.

The measurement-setting choices also do not fully realize the independence condition required for a loophole-free LF experiment. The relevant requirement is not certified true randomness or perfectly uniform setting statistics, but independence of the setting choices from physical variables that may otherwise influence the experimental outcomes. Here, the settings are generated by measuring choice qubits on the same quantum processor that implements the EWFS circuit. A stronger implementation would use externally generated random setting choices supplied to the quantum processor as classical inputs during the experiment.

The EWFS used here is minimal in the sense that it contains one friend, Charlie, and two superobservers, Alice and Bob, rather than two friends as in the original two-friend scenario. This is sufficient for the LF inequality considered here, since the scenario admits LF violations already with one friend. Avoiding a second friend is convenient for near-term hardware, because the dominant practical limitation is noise and the noise generally increases with circuit size.

\section{\label{sec:conclusion}Conclusion}

This work investigated a new way of implementing the ``friend'' in a Local Friendliness experiment. Simple agent-like observers were implemented as reversible quantum circuits and embedded in a one-friend Extended Wigner's Friend Scenario. The agents store measurement outcomes, condition later operations on stored information, and, in the case of the Betting Agent, use Born-rule probabilities to choose an action.

Ideal simulations reproduce the theoretical maximal violation for all implemented agents. Noise simulations yield smaller violations, while hardware runs on \texttt{ibm\_marrakesh} show positive LF violations for all four agents. The observed violations also remain above the estimated relaxed LF bounds within the assumptions of the applied estimation method.

The implemented agents should not be taken as satisfying the assumptions of Thoughtful Local Friendliness. Rather, they provide simple intermediate examples between elementary qubit-like friends and the much more sophisticated artificial observers envisioned in long-term proposals. A natural next step is therefore to move towards larger and more structured reversible agents, potentially combining agent functionality with increased branch complexity, while improving quantum hardware and ultimately the loophole structure of the experiment.

Although the resulting agent systems studied in this work remain far from human-level AI, they form a first step from qubit-like friends towards those more demanding observer models. The most sophisticated agent presented here is the Betting Agent. Its policy is motivated by expected utility and compiled into a fixed condition--action rule. It provides a minimal example of acting according to a policy derived from quantum-theoretic probabilities, while remaining quantum-coherent -- a rudimentary ``quantum user of quantum mechanics''. 

\bigskip

\section*{Acknowledgements}

The code developed for this work builds upon earlier code developed by Will Zeng and collaborators for Ref.~\cite{zeng_towards_2025}. We acknowledge Will Zeng for useful discussions and assistance with adapting that code for the present work. We also acknowledge discussions with and feedback from Howard Wiseman, Renato Renner and Marwan Haddara. EGC acknowledges funding from the Australian Research Council Discovery project DP250102162. OpenAI Codex (GPT-5 and GPT-6) was used to assist with code development and debugging, and for checking for typos and errors prior to submission. All AI-assisted code and its outputs were reviewed and verified. 

\section*{Data and Code Availability}

The code used for this work is available on GitHub at \url{https://github.com/lauxj/agent-like-observers-ewfs}, including project notebooks to follow the procedure. New experimental runs can be performed and all results presented in this work, as well as plots, can be reproduced from the repository.

\bibliography{references}% Produces the bibliography via BibTeX.

%\clearpage
% \onecolumngrid
% \par\bigskip
% \noindent\rule{\textwidth}{0.4pt}
% \par\bigskip
% \twocolumngrid

\begingroup
  \renewcommand{\addcontentsline}[3]{} % Avoid duplicate References entry
  \bibsection
\endgroup

\appendix

\section{\label{app:lf_details}Derivation of the Optimal Angles}

In this appendix we justify the measurement angles used for the implementation of the LF inequality in the one-friend Extended Wigner's Friend Scenario. The relevant qubits are Alice's system qubit \(S_A\), Bob's system qubit \(S_B\), and Charlie's memory qubit \(M\).

The circuit starts with the Bell state
\begin{equation}
    \ket{\Phi^+}_{S_A S_B}
    =
    \frac{1}{\sqrt{2}}
    \left(
        \ket{00}_{S_A S_B}+\ket{11}_{S_A S_B}
    \right).
\end{equation}
Charlie's measurement of \(S_A\) is implemented coherently by a CNOT gate from \(S_A\) to \(M\), which gives
\begin{equation}
    \ket{\mathrm{GHZ}}_{S_A M S_B}
    =
    \frac{1}{\sqrt{2}}
    \left(
        \ket{000}_{S_A M S_B}+\ket{111}_{S_A M S_B}
    \right).
\end{equation}

For Alice's first setting, Alice asks Charlie for their outcome, which is stored in the memory qubit \(M\). Since this outcome is encoded in the computational basis, Alice's readout corresponds to a \(\sigma_z\) measurement on \(M\), giving
\begin{equation}
    A_1=\sigma_z^{(M)}.
\end{equation}
For Alice's second setting, Charlie's interaction is undone and Alice applies \(R_y(\alpha)\) before measuring \(S_A\). With
\(R_y(\theta)=\exp(-i\theta\sigma_y/2)\),
\begin{equation}
    R_y(\theta)^\dagger \sigma_z R_y(\theta)
    =
    \cos\theta\,\sigma_z-\sin\theta\,\sigma_x,
\end{equation}
and therefore
\begin{equation}
    A_2
    =
    \cos\alpha\,\sigma_z^{(S_A)}
    -
    \sin\alpha\,\sigma_x^{(S_A)}.
\end{equation}
Bob's observables are analogously
\begin{equation}
    B_j
    =
    \cos\beta_j\,\sigma_z^{(S_B)}
    -
    \sin\beta_j\,\sigma_x^{(S_B)},
    \qquad j=1,2.
\end{equation}

For the setting in which Alice asks Charlie for their outcome,
\begin{equation}
    \langle A_1 B_j\rangle=\cos\beta_j.
\end{equation}
When Alice reverses Charlie's interaction, the relevant state is again
\(\ket{\Phi^+}_{S_A S_B}\), for which
\begin{equation}
    \langle A_2 B_j\rangle
    =
    \cos\alpha\cos\beta_j+\sin\alpha\sin\beta_j
    =
    \cos(\alpha-\beta_j).
\end{equation}

The LF expression is therefore
\begin{equation}
\begin{split}
    S_{\mathrm{LF}}(\alpha,\beta_1,\beta_2)
    ={}&
    -\cos\beta_1+\cos\beta_2 \\
    &-\cos(\alpha-\beta_1)
    -\cos(\alpha-\beta_2)-2.
\end{split}
\end{equation}

Defining
\begin{equation}
    u=\frac{\beta_1+\beta_2}{2},
    \qquad
    v=\frac{\beta_2-\beta_1}{2},
\end{equation}
gives
\begin{equation}
    S_{\mathrm{LF}}
    =
    -2\sin u\sin v
    -
    2\cos(\alpha-u)\cos v
    -
    2.
\end{equation}
The expression satisfies
\begin{equation}
\begin{aligned}
S_{\mathrm{LF}}+2
&\leq 2|\sin v|+2|\cos v|\\
&\leq 2\sqrt{2}.
\end{aligned}
\end{equation}
Equality is attained when
\begin{equation}
\begin{aligned}
|\sin v|=|\cos v|&=\frac{1}{\sqrt{2}},\\
\sin u\,\sin v&=-\frac{1}{\sqrt{2}},\\
\cos(\alpha-u)\cos v&=-\frac{1}{\sqrt{2}}.
\end{aligned}
\end{equation}
One convenient choice is
\begin{equation}
    u=\frac{\pi}{2},
    \qquad
    v=-\frac{\pi}{4},
    \qquad
    \alpha=\frac{3\pi}{2},
\end{equation}
which gives
\begin{equation}
    \beta_1=\frac{3\pi}{4},
    \qquad
    \beta_2=\frac{\pi}{4}.
\end{equation}

For these angles,
\begin{equation}
\label{eq:correlators_max_value}
\begin{aligned}
    \langle A_1B_1\rangle &= -\frac{1}{\sqrt{2}},&
    \langle A_1B_2\rangle &=  \frac{1}{\sqrt{2}},\\
    \langle A_2B_1\rangle &= -\frac{1}{\sqrt{2}},&
    \langle A_2B_2\rangle &= -\frac{1}{\sqrt{2}}.
\end{aligned}
\end{equation}
Thus
\begin{equation}
    S_{\mathrm{LF}}^{\max}
    =
    2\sqrt{2}-2
    \approx0.828.
\end{equation}

\section{\label{app:hardware_implementation}Hardware Implementation Details}

\begin{table*}[t!]
\caption{Logical-to-physical qubit mappings used for the
\texttt{ibm\_marrakesh} hardware runs.}
\label{tab:qubit_mapping}
\begin{ruledtabular}
\begin{tabular}{lcc}
Agent & Logical qubit order & Physical qubits \\
\hline
Reflex &
$(S_B,S_A,M,R,A_C,B_C)$ &
$(10,11,12,13,0,155)$ \\
Guessing &
$(S_B,S_A,M_1,M_2,G,A_C,B_C)$ &
$(10,11,12,18,13,0,155)$ \\
Always-$3/4$ &
$(S_B,S_A,M_1,M_2,W_0,W_1,A_C,B_C)$ &
$(18,11,12,10,13,9,0,155)$ \\
Betting &
$(S_B,S_A,M_1,M_2,W_0,W_1,A_C,B_C)$ &
$(18,11,12,10,13,9,0,155)$
\end{tabular}
\end{ruledtabular}
\end{table*}

\subsection{Qubit Layout}

The qubits of \texttt{ibm\_marrakesh} are connected by a fixed coupling map, so that a two-qubit gate can only be applied directly between connected qubits. If two qubits that need to interact are not directly connected, additional routing operations are required. Since these operations increase the circuit depth and therefore the sensitivity to noise, the circuits in this work are constructed such that no SWAP gates are needed.

Each agent has a different pattern of required two-qubit interactions. The logical qubits are therefore mapped to physical qubits such that all required connections are directly available.

Using IBM calibration data for the main error sources, such as readout errors, CZ errors, and coherence times, a low-noise qubit placement was chosen. 

The resulting logical-to-physical qubit mappings used for the hardware runs are listed in Table~\ref{tab:qubit_mapping}. The choice qubits \(A_C\) and \(B_C\) were mapped to physical qubits \(q_0\) and \(q_{155}\), respectively, for all agent implementations. Since the choice qubits do not participate in two-qubit gates with the EWFS system, their physical placement is largely independent of the connectivity requirements of the agent circuits.

\subsection{Native Gate Decomposition}

The gate basis used for transpilation to \texttt{ibm\_marrakesh} in this work is 

\begin{equation}
\{\mathrm{id},R_Z(\theta),SX,X,CZ\}. 
\end{equation}
The logical gates appearing in the agent circuits are translated into this native gate set as shown in Table~\ref{tab:gate_translation}.

\begin{table}[tbp]
\caption{Gate translation of the logical gates used in the agent circuits into the native gate set of the Heron-r2 architecture, up to a global phase. Decompositions are listed in left-to-right execution order.}
\label{tab:gate_translation}
\begin{ruledtabular}
\begin{tabular}{lc}
Logical gate & Native decomposition \\
\hline
$R_Y(\theta)$ &
$R_Z(0),SX,R_Z(\theta+\pi),SX,R_Z(3\pi)$ \\
$H$ &
$R_Z(\pi/2)\,SX\,R_Z(\pi/2)$ \\
$X$ & $X$ \\
$\mathrm{CNOT}_{c\rightarrow t}$ &
$(I_c\otimes H_t)\,CZ_{c,t}\,(I_c\otimes H_t)$ \\
\end{tabular}
\end{ruledtabular}
\end{table}

The circuits were transpiled using Qiskit's integrated \texttt{transpile} function with \texttt{optimization\_level=0}, which avoids additional optimization changes and preserves the order and logic of the original circuit. The resulting circuits are expressed in the native gate set and physical-qubit layout of \texttt{ibm\_marrakesh} and represent the circuits executed in the experiment.

\section{\label{app:transmission_circuits}Relaxed LF Estimation}

Relaxed LF allows Alice's \(x=1\) outcome to differ from the agent's recorded outcome with probability \(\varepsilon\). To estimate this parameter experimentally, one needs an estimate of how accurately the value assigned to \(c\) is later recovered in Alice's readout \(a\).

For this estimate, the initial measurement interaction $\mathrm{CNOT}_{S_A\rightarrow M}$, or $\mathrm{CNOT}_{S_A\rightarrow M_1}$ for the agents with two memory registers, is replaced by a fixed preparation of the corresponding memory qubit. For $c=0$, this qubit is left in its initial state $\ket{0}$, while for $c=1$, an $X$ gate prepares it in $\ket{1}$. The subsequent agent operations and Bob's setting-dependent operations are retained, whereas Alice's classically conditioned $x=2$ block is omitted. An explicit delay of $83\,dt$, $96\,dt$, $121\,dt$, or $115\,dt$ is inserted on the first memory qubit immediately before readout for the Reflex, Guessing, Always-$3/4$, and Betting agents, respectively, where $dt$ is the time unit used by the circuit's delay instruction. Only shots with $x=1$ are included. Comparing Alice's final readout with the intended preparation label gives an estimate of the memory-transmission error in this modified circuit.

The corresponding procedure is applied separately to all implemented agents. The resulting estimate is used to evaluate the relaxed LF bound discussed in Sec.~\ref{sec:relaxed_lf_results}. As emphasized in the main text, this procedure provides an estimate of \(\varepsilon\), rather than a rigorous upper bound. The calibration circuits omit the entangling interaction that forms the first memory record and therefore do not characterize errors or crosstalk associated with that interaction. Replacing this interaction also changes the correlations between the memory and the remaining registers. Furthermore, omitting Alice's conditional block can change the hardware execution schedule even for $x=1$, and the inserted delay does not establish equivalent noise exposure. Transferring the calibration estimates to the original EWFS experiment therefore requires an additional assumption; these systematic differences are not included in the reported statistical uncertainties.

For each agent, both memory preparations are tested over 10 runs, using \(10^3\) shots per preparation and per run on hardware, \(10^4\) in the noise simulation, and \(10^5\) in the noiseless simulation. After conditioning on \(x=1\), the memory-readout error frequencies are averaged over runs separately for each preparation and each setting \(y\). The two preparation means are then combined with equal weight, and \(\varepsilon_{\mathrm{est}}\) is the larger of the resulting estimates for \(y=1\) and \(y=2\). Thus, averaging over runs precedes maximization. Equal weighting assumes equally probable memory outcomes in the ideal experiment. For each setting, the uncertainty is obtained by combining the two preparation-specific standard errors of the mean in quadrature and dividing by two, assuming independent preparation estimates. The errors shown in Table~\ref{tab:epsilon_comparison} and Fig.~\ref{fig:tracking_values} are those for the selected setting; they do not account for uncertainty associated with selecting the maximum. Since preparation accuracy and the transfer of calibration errors to the original experiment are not independently certified, this procedure provides a practical diagnostic rather than a rigorous upper bound.

%\clearpage

\end{document}